# Quran Recitation Recognition using End-to-End Deep Learning

Ahmad Al Harere*, Khloud Al Jallad

Department of Information and Communication Engineering,
Arab International University, Daraa, Syria.

*Corresponding author. E-mail(s): 201710632@aiu.edu.sy Contributing authors: k-aljallad@aiu.edu.sy

**Abstract**

The Quran is the holy scripture of Islam, and its recitation is an important aspect of the religion. Recognizing the recitation of the Holy Quran automatically is a challenging task due to its unique rules that are not applied in normal speaking speeches. A lot of research has been done in this domain, but previous works have detected recitation errors as a classification task or used traditional automatic speech recognition (ASR). In this paper, we proposed a novel end-to-end deep learning model for recognizing the recitation of the Holy Quran. The proposed model is a CNN-Bidirectional GRU encoder that uses CTC as an objective function, and a character-based decoder which is a beam search decoder. Moreover, all previous works were done on small private datasets consisting of short verses and a few chapters of the Holy Quran. As a result of using private datasets, no comparisons were done. To overcome this issue, we used a public dataset that has recently been published (Ar-DAD) and contains about 37 chapters that were recited by 30 reciters, with different recitation speeds and different types of pronunciation rules. The proposed model performance was evaluated using the most common evaluation metrics in speech recognition, word error rate (WER), and character error rate (CER). The results were 8.34% WER and 2.42% CER. We hope this research will be a baseline for comparisons with future research on this public new dataset (Ar-DAD).

**Keywords:** Deep Learning, End-to-End, Speech Recognition, Quran recitation, Natural Language Processing

## 1. Introduction

Speech communication is an important way of social interaction to convey our thoughts, ideas, and emotions to others. Moreover, speech is also a crucial tool for learning and education, as it is the primary way in which information is exchanged between teachers and students.

Processing speech using computers and artificial intelligence is a complex task that has been a hot research topic in recent years. One of the main challenges is automatically transcribing spoken words into text. The performance of such systems has greatly improved in recent years due to advancements in deep learning.

Arabic language is a rich language with a long history and cultural significance. It is spoken by over 400 million people worldwide, and it is the official language in many countries. Arabic language recognition is a challenging task because of a lack of resources and, having many variations in pronunciation and dialects. Although these difficulties, researchers have made significant progress in developing Arabic speech recognition systems, which can be used in tasks such as automatic speech transcription and translation, and also in speech-enabled applications such as voice assistants and chatbots.

The Holy Quran, which was revealed in Arabic language, holds a central place in the hearts and minds of Muslims. Quran is considered the holy book of Islam and the words of Allah. In addition, it is a guide for all aspects of life, providing moral and spiritual teachings and a source of inspiration and guidance. The Holy Quran recitation recognition is a particularly challenging task because of the specific requirements of the task, such as recognizing different recitation styles and checking the correct pronunciation of Tajweed rules, a set of pronunciation rules that must be applied to recite the Quran in the same way that the Prophet Muhammad did. Also, Quran recitation includes many unique sounds and intonations that are not used in other forms of Arabic spoken language. Many researches have been done on Quran recitation processing over time to make Quran recitation easier and more accessible to a wider audience. One of these hot topics of researches is Quran recitation recognition systems for tasks such as recognizing reciters and detecting errors in recitation based on Tajweed rules.

Although the research topic of recognizing the recitation of the Holy Quran has been a hot topic in recent years, most of research papers are limited to detecting mispronunciation of words or some Tajweed rules on private small datasets. Some researchers proposed detecting the mispronunciation of reciting some verses directly from speech features, while others proposed converting the recitation into text using traditional ASR. As all previous works were conducted on small private data sets of few chapters, they do not cover a large number of examples of Tajweed rules and different speeds of recitation. As a result, previous works are still not effective to be applied in real-life applications because detecting only the mispronunciation of the Tajweed rules is less important than detecting the mispronunciation of words. Moreover, the use of traditional ASR in recitation recognition suffers from many problems as ASR models require specific forms of datasets which are not available for Quran recitations yet. We will discuss those problems in detail later in the related works section.

This paper aims to fill these gaps through the use of end-to-end deep learning methodologies that overcome the problems of using traditional ASR. Experiments were done on the Ar-DAD dataset, which is a large public dataset covers most of Tajweed rules and the different speeds of recitation through the participation of about 30 readers in reciting the verses.

The main contributions of this work are:

- Using the end-to-end methodology instead of the traditional ASR, for automatic phoneme alignment, so we do not need alignment tools anymore. To the best of our knowledge, the task of Quran recitation recognition has not yet been tackled using an end-to-end deep learning approach, and this work is to fill this gap.
- We have evaluated our model on a big public dataset, so it can be a baseline model for comparisons later.
- By comparing the predicted text with the real text, our solution can determine the type of error (deleting, substituting, adding) at the level of words, characters, and some Tajweed rules, with the exact position of the error.

In terms of limitations, first, all the samples in the Ar-DAD dataset are for men reciters, which makes the model less robust for recognizing the recitations by women and children. Second, the Ar-DAD dataset contains samples from only one recitation form, the 'Hafs from Aasim' recitation [1][2]. However, recitation of the Holy Quran has ten recitation forms (Qira'at) approved by scholars [3]. The differences between these ten forms are mainly in the pronunciation of certain words, prolongation, and intonation [1]. Each form of recitation has its own unique features that distinguish it from the others. This may lead the model to incorrectly recognize samples with different forms of recitation.

## 2. Related Works

The Arabic language has several forms, two formal forms, and many slang forms. As for formal forms, Arabic has the classical Arabic language (CA), which is the language of the Holy Quran, and the modern standard Arabic (MSA) that is used in, news, books, … etc. As for the slang forms, Arabic has many dialects that differ from one country to another. Since this paper is for the recognition of classical Arabic speech, the literature review will focus only on papers done for the recognition of the Holy Quran recitation. But there are no researches done using end-to-end on Holy Quran, so we discussed some researches that used end-to-end deep learning methodologies on Modern Standard Arabic.

There are several challenges in Quran Recitation Recognition. First, there are several letters In the Arabic language that have confusing pronunciation, as they share the articulation way out and some characteristics.[4], [5]. This confusion significantly impacts speech recognition models, as it leads to errors in the recognition of Arabic speech and decreases the overall accuracy of the model. For example, suppose a model is not able to distinguish between the letters "ص" (ṣād) and "س" (seen). In that case, it may incorrectly recognize the word "الصراط" (alserat) as "السراط" (alserat) which can lead to misinterpretation of the text and an increase in both Word and Character Error Rates. Second, the complex and nuanced rules of Tajweed make it difficult for an AI model to accurately recognize recitation because some of these rules change the pronunciation way of letters when applied. For example, the Turning rule (الإقلاب) when applied, the pronunciation of "ن" (noon) becomes "م" (meem). Third, having different forms of Quran Recitation make the ASR more complex task because of different ways of pronouncing some of the Tajweed rules, and the length or manner of pronunciation may vary depending on the context and the reciter. For example, Separated Lengthening (المد المنفصل), which is the prolongation of a letter that comes at the end of a word, can be pronounced for 2, 4, or 5 counts in length. Another example is Concealment (الإخفاء), which is the rule of hiding the pronunciation of certain letters, it can be pronounced in different degrees of hiddenness.

Moreover, one of the main difficulties in the Quran recitation detection task is that there are three different speeds for reciting the Quran: Hadr (حدر), Tahqeeq (تحقيق), and Tadweer (تدوير) [6]. Each speed has its own unique advantages, and each of them is used to help listeners understand the Quran better and to get the most out of the recitation.

Hadr (حدر): is typically considered the fastest speed of recitation, where the emphasis is on fluency and the ability to recite large portions of the Quran quickly and smoothly. This speed is particularly useful for those who are already familiar with the Quran, have a deep understanding of the Quran, and are proficient in the rules of Tajweed.

Tadweer (تدوير): is the moderate speed of recitation, where the emphasis is on proper pronunciation and intonation, while still maintaining a relatively moderate pace. This speed is particularly useful for those who have a basic knowledge of Tajweed and are trying to improve their recitation and pronunciation.

Tahqeeq (تحقيق): is the slowest speed of recitation, where each letter is pronounced clearly and deliberately, allowing the listener to fully understand the meaning of each verse. This speed is particularly useful for those who are learning to recite the Quran for the first time or for those who are still not familiar with the rules of Tajweed.

In this section, we will discuss two basic types of researches, researches on Quran and researches that used end-to-end deep learning on Modern Standard Arabic (MSA) as there are no end-to-end experiments on Quran.

As for researches on Quran, we have two basic types of methodologies.

- Researches based on detecting mispronunciation from speech directly. Either for Tajweed Rules Mispronunciation or Character Mispronunciation. Table 1 shows a comparison of these studies.
- Researches based on traditional ASR that convert speech to text, then detect mispronunciation by comparing the result text with Quran text.

## 2.1 Researches based on detecting mispronunciation from speech directly

Researches based on detecting mispronunciation from speech directly were done using traditional methodologies, such as Hidden Markov Model (HMM) and Gaussian Mixture Model (GMM), or machine learning models such as Support Vector Machine (SVM) and Multi-Layer Perceptron (MLP), as follows:

Hassan et al. [7] developed a solution to recognize Qalqalah Kubra[1] (القلقلة الكبرى) pronunciation using Multilayer Perceptron as a classifier and MFCC as features extraction. The dataset used contains 50 samples, each with correct and incorrect pronunciation, and the achieved results ranged from 95% to 100%.

Al-Ayyoub et al. [8] used machine learning to build a model for the automatic recognition of Quran Recitation Rules (Tajweed). This model was able to determine the recitation correctness of the following eight rules of intonation: EdgamMeem[2], EkhfaaMeem[3], Tafkheem Lam[4], Tarqeeq Lam[5], Edgam Noon[6](Noon), Edgam Noon (Meem), Edgam Noon (Waw) and Edgam Noon (Ya'). The authors used a dataset that consists of 3,071 audio files, each containing a recording of exactly one of the eight rules under consideration (in either the correct or the incorrect usage of the rule). For feature extraction, many feature extraction techniques were used such as Linear Predictive Code (LPC), Mel-frequency, Cepstral Coefficients (MFCC), Multi-Signal Wavelet Packet Decomposition (WPD), and Convolutional Restricted Boltzmann Machines (CRBM). As for classification, several classifiers were used such as k-Nearest Neighbors (KNN), Support Vector Machines (SVM), Artificial Neural Networks (ANN), and Random Forest (RF), with accuracy of 96% using SVM.

Alagrami et al. in [9] proposed a solution that makes use of threshold scoring and support vector machine (SVM) to automatically recognize four different Tajweed rules (Edgham Meem, Ekhfaa Meem, takhfeef Lam, Tarqeeq Lam) with 99% accuracy, where the filter banks were adopted as feature extraction. The dataset used contained about 657 records of Arabic natives and non-natives, each rule has 160 records, and each of them is either the correct pronunciation or the wrong pronunciation of this rule.

A Tajweed classification model was developed by Ahmad et al. in [10]. This solution focused on a set of Tajweed rules called "the Noon Sakinah rules" and in particular the rule of "Idgham" with and without "Ghunnah". Mel-Frequency Cepstral Coefficient and a neural network were used for the feature extraction and the classification process, where Gradient Descent with Momentum, Resilient Backpropagation, and the Levenberg-Marquardt optimization algorithms, were used to train the neural network. The Levernberg Marquardt algorithm achieved the highest test accuracy (77.7%), followed by Gradient Descent with Momentum (76.7%) and Resilient Backpropagation (73.3%). The dataset used was 300 audio files of recitation of two famous reciters, and each is a recitation of one of those Tajweed rules.

Nahar et al. in [11] took a different path as they proposed a recognition model to recognize the "Qira'ah" from the Holy Quran recitation precisely 96%, since according to the narration "hadith" No. 5041, taken from [12], the Holy Quran has seven main reading modes, known as "Qira'at," which are acknowledged as the most popular methods of reciting from the Holy Quran, and three complementary readings of the seven. This model used the Mel-Frequency Cepstrum Coefficients (MFCC) features and Support Vector Machine (SVM), where the authors have built a dataset has 10 categories, each one representing a type of Holy Quran recitation or "Qira'ah", with a total of 258 wave files.

For detecting letter and word errors from speech directly, the researches carried out in this field used classifiers trained on datasets containing samples of mispronunciation and correct pronunciation of specific verses, or they stored the characteristics of correct recitation of certain verses in a database. Compare users' recitations with stored recitations and then calculate the similarity using a threshold. One of the most significant shortcomings of this methodology is that it can only categorize the recitations of the verses that were presented in the datasets.

---

[1] Qalqalah: it is the vibration of sound at the end of the pronunciation of a letter.
The letters of Qalqalah: قطب جد.

[2] EdgamMeem :If the start of the word begins with a meem and is followed with a meem sakinah, then merge the words through the meem and apply ghunnah.

[3] EkhfaaMeem: If a [ب] is followed after a meem sakinah, then apply a ghunnah while hiding the meem sakin before continuing to the [ب].

[4] Tafkheem Lam: If there is a Fatha or a Dhamma before the word of Allah or Allahum, then laam in Allah will be heavy.

[5]Tarqeeq Lam: If there is a kasrah (ِ) before the word Allah, then the Laam in Allah or Allahum will be light.

[6]Edgam Noon: If the Noon Sakin (ْ) or the Tanween is followed by any of Ingham letters, the reciter should skip the Noon or the Tanween and pronounce the following letter.

One of the earliest works done in this field is the work of Tabbal et al. in [13], where an automated verse delimiter and an error detection model were developed for the recitation of the Holy Quran. HMM classifier and Mel Frequency Cepstral Coefficient (MFCC) features were used on a private dataset that is one hour of recitations of Surah Al-Ikhlas only. The best accuracy obtained by this solution was 85% for females and 90% for males.

Putra et al. [14] developed software for Quranic Verse Recitation Learning. The solution proposed used Mel Frequency Cepstral Coefficient (MFCC) for feature extraction and the GMM model as a classifier. In order to test the reliability and accuracy of correction, a data set was collected from ten speakers reading some verses incorrectly and correctly for each of them. The achieved correction accuracy was 90% for hija'iyah letters (Arabic Alphabet Letters) pronunciation, 70% for recitation law where the law might be idgham, ikhfa' or idhar, and 60% for the combination of pronunciation and recitation law.

In [15], Rahman et al. proposed an automated checking system to learn children the correct recitation of the Holy Quran. Mel-Frequency Cepstral Coefficient (MFCC) was used for feature extraction, and Hidden Markov Model (HMM) was used for classification and recognition. Using the HMM algorithm, the model can identify and highlight any discrepancy or inconsistency in children's recitation by comparing it with the correct teacher's recitation that was stored in a database, where only one chapter of the Quran was supported, Surah Al-Fatiha.

Muhammad et al. [16] proposed E-Hafiz system to facilitate reciting learning of the holy Quran, where Mel Frequency Cepstral Coefficient (MFCC), Vector Quantization (VQ), and Calculation of distance between vectors were used to extract the features, reduce the number of features vectors and compare the result with the threshold value. A dataset of 10 expert recitations of the first 5 surahs of the Holy Quran was used, and recognition accuracy of 92%, 90%, and 86% was achieved for men, children, and women, respectively.

Rajagede and Hastuti [17] propose a model to help users to check their Al-Quran memorization using the Siamese Long Short-Term Memory (LSTM) Network. Siamese LSTM network was used to check the similarity between two samples, so it verifies the recitation by matching the input with existing data for a verse read, without performing a speech-to-text extraction process. Two Siamese LSTM architectures were compared: the Siamese-Classifier, which employed binary classification, and the Manhattan LSTM, which produced a single numerical value to indicate similarity. In addition, the performance of the models was compared with Mel Frequency Cepstral Coefficient (MFCC), Mel Frequency Spectrum Coefficient (MFSC), and Delta Features, where an F1-score of 77.35% was given by using MFCC with delta features and Manhattan LSTM, as the best result obtained. Four reciters who recited 48 verses from the last 10 Surahs of the Quran provided the data used to train the model.

## 2.2 Researches based on traditional ASR

Very few researches suggested converting the recitation speech into text by using Automatic Speech Recognition techniques, then detecting mispronunciation by comparing the result text with Quran text. This methodology is better than models that detect mispronunciation from speech directly, as it helps to detect various errors in the user's recitation by comparing the predicted text with the original text of the verses. Moreover, no need for wrong pronunciation samples.

To detect mistakes made during the recitation of the Holy Quran, Tabbaa and Soudan [18] created a computer-aided recitation training solution that combined Automatic Speech Recognition (ASR) with a classifier-based approach to increase the detection rate. This solution detects errors in two phases: the HMM-based ASR recognizes the recitation, and then the classifier applies, where two classifiers were used, one to distinguish between the stressed and non-emphasized pronunciations of the Arabic letter "R", and the other to separate closely related and frequently mixed-up letter pronunciations. The HMM recognizer was trained using CMU Sphinx, and the classifiers were built using WEKA (Waikato Environment for Knowledge Analysis), where numerous machine learning algorithms were tested. Up to 7 hours of recitations were recorded by phone calls to a TV program, the recitation scholar reads a page from the Quran before listening to the students' recitations and correcting any mistakes. According to the results, the system has a word-level accuracy of 91.2%, where it has been tested on 60 minutes of continuous recitation.

Al-Bakeri et al. in [19] introduced an ASR integrated with a self-learning environment that depends on MVC architectures to correct recitation automatically. The speech recognition model was built using open-source CMU Sphinx tools, which also contain the Hidden Markov Model (HMM) code that was chosen for feature extraction, feature training, and pattern recognition. Also, language models were used in the process of building the system and built using CMU-CSLMT tools [20]. The corpus contains the recitations of two short chapters, Surah Al-Ikhlas and Surah Al-Rahman, which were recorded by 10 famous Quran reciters. To assess the ASR performance, the word error rate (WER) was used, where the ASR output was compared with the correct words of the verse considering insertions, deletions, and substitution, so the correctness was presented as a number ranged between 47.47% and 75.2% as reported.

A speech recognizer for the Holy Quran was introduced by Tantawi et al. in [21]. This solution is able to recognize the recitation of some verses in addition to some Tajweed rules that were taken into account during the development process, where it was trained using 32 recordings of Chapter 20 from the Holy Quran according to Narration of "Hafs on the authority of Asim" (One of the ten reading forms of the Quran). The pronunciation dictionary for the Holy Quran recitations was built using an automated tool proposed by [22], where the transcription was passed to it to build the dictionary. As for the language model, The SRI Language Modeling (SRILM) toolkit [23] was used. With the KALDI toolkit, numerous experimental configurations with various dataset sizes and Tajweed rules were used. The best experimental setup used MFCC features and Time Delay Neural Networks (TDNN), where Word Error Rates (WER) and Sentence Error Rates (SER) ranged from 0.27 to 6.31% and 0.4 to 17.39%, respectively.

However, this methodology is not effective in recognizing Quran recitation as it is based on traditional ASR, because there was a problem with the alignment process needed to train the acoustic model. The traditional ASR consists mainly of three models:

- Pronunciation dictionary that converts words from the original language into a series of phonemes that express the pronunciation of these words.
- Acoustic model that connects phonemes with the features extracted from the corresponding sound.
- Language model that is responsible for determining the most likely sequence of words based on the context and grammar of the language.

Figure 1 shows how these components together:

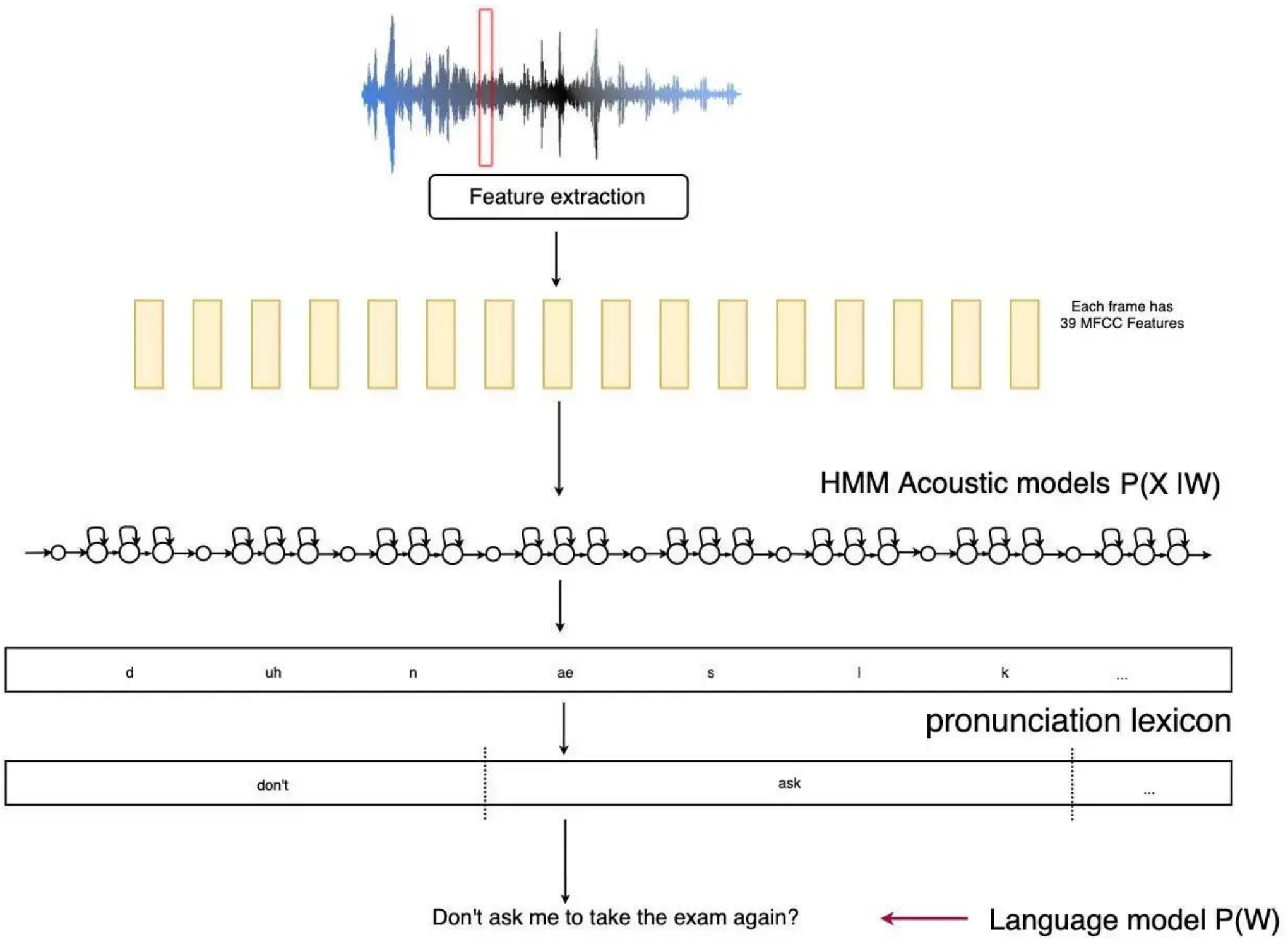

**Fig. 1** Traditional ASR workflow [24]

Training datasets must contain the correct alignment between the acoustic frames and phonemes in order to train the acoustic models. This is one of the biggest problems in this field, as no dataset of this format is available for Quran recitations, unlike the MSA, which has several datasets for this format, such as the KAPD dataset [25] and the Nawar Halabi dataset [26]. As a result, researchers who proposed this method used automated tools that perform this alignment process, but the results were not good enough to recognize the recitation efficiently.

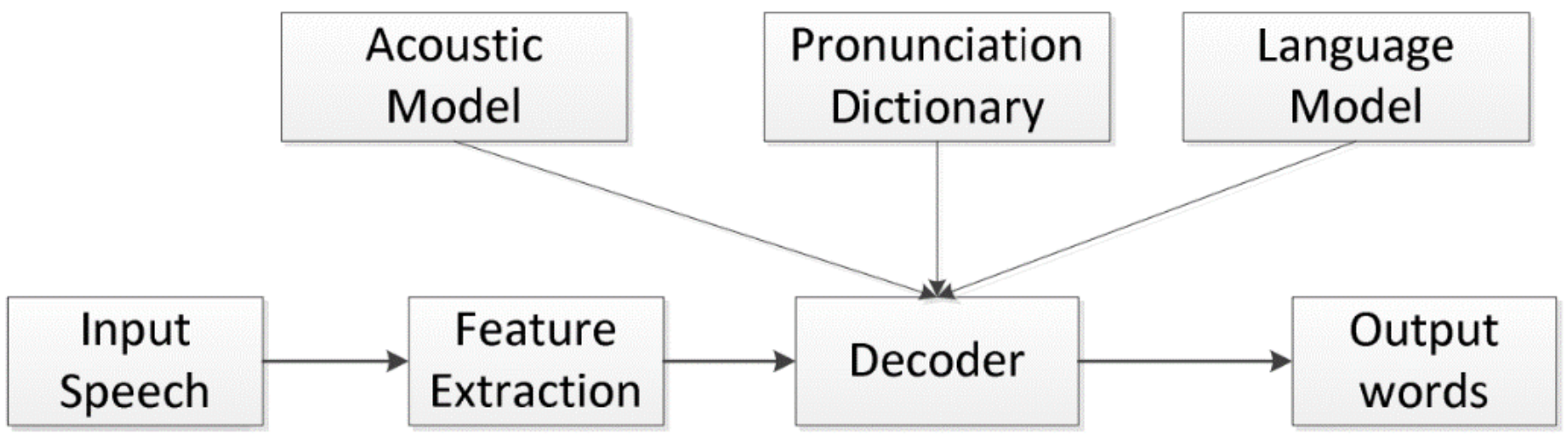

**Fig. 2** Traditional ASR pipeline [27]

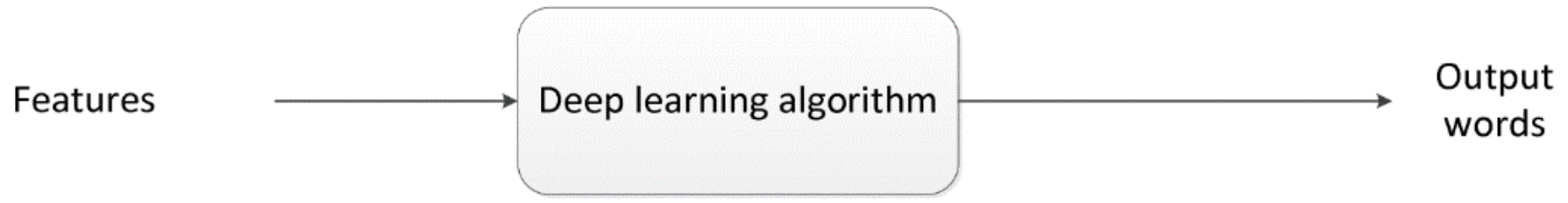

**Fig. 3** End-to-End ASR pipeline [27]

For this reason, we proposed using the end-to-end methodology instead of the traditional ASR. The end-to-end models can do the alignment process automatically without any need for additional tools, and convert acoustic features to text transcription directly without the need for all other components needed in traditional ASR, which makes them more efficient and suitable for Quran recitation recognition. Figure 2 and Figure 3 show a comparison between the conventional ASR pipeline and the end-to-end pipeline. To the best of our knowledge, the task of Quran recitation recognition has not yet been tackled using an end-to-end deep learning approach, and this work is to fill this gap.

**Table 1** A comparison of some recitation recognition works

| Work | Dataset | Feature Extraction | Model | Recognition Level | Accuracy |
|---|---|---|---|---|---|
| [7] | 100 samples for Qalqalah pronunciation | Mel-Frequency Cepstral Coefficient (MFCC) | Multi-Layer Perceptron (MLP) neural network | One Tajweed rule (Qalqalah) | 95% - 100% |
| [8] | 3071 audio files, each containing a recording of exactly one of the eight rules | CRBM, MFCC, WPD, HMM-SPL | Support Vector Machine (SVM) | Eight Tajweed rules (EdgamMeem,EkhfaaMeem, Tafkheem Lam, Tarqeeq Lam, Edgam Noon (Noon), Edgam Noon(Meem), Edgam Noon(Waw) and Edgam Noon(Ya')) | 96.4% |
| [9] | 657 records of Arabic natives and non-natives | Filter Banks | Support Vector Machine (SVM) | Four different Tajweed rules (Edgham Meem, Ekhfaa Meem, takhfeef Lam, Tarqeeq Lam) | 99% |
| [10] | 300 audio files of recitation of two famous reciters | Mel-Frequency Cepstral Coefficient (MFCC) | Neural Network | "Idgham" rules with and without "Ghunnah" | 73.3% - 77.7% |
| [11] | 258 wave files | Mel-Frequency Cepstral Coefficient (MFCC) | Support Vector Machine (SVM) | Classification of recitation into one of ten recitations types ("Qira'at") | 96% |
| [14] | Voice recorded from an expert | Mel-Frequency Cepstral Coefficient (MFCC) | Gaussian Mixture Model (GMM) | Letter level and some Tajweed rules | Letters: 90%<br>Tajweed rules: 70%<br>Combination: 60% |
| [16] | Consists of 10 expert recitations of the first 5 surahs of the Holy Quran. | Mel-Frequency Cepstral Coefficient (MFCC) | Threshold based on Euclidean distance | Word level for verses in the dataset | 86% - 92% |
| [13] | About 1 hour of audio recitations of Sourate Al-Ikhlass | Mel-Frequency Cepstral Coefficient (MFCC) | Hidden Markov Model (HMM) based on Sphinx | Sourate Al-Ikhlass with the most important Tajweed rules | 85% - 90% |
| [28] | Some specific verses recited by experts | Mel-Frequency Cepstral Coefficient (MFCC) | Threshold based on distance | Word level for verses in the dataset | 90% - 92% |
| [17] | 48 verses from the last 10 Surahs recited by four reciters | Mel-Frequency Cepstral Coefficient (MFCC) | Siamese LSTM | 48 verses from the last 10 Surahs | 77.35% |

As our proposed model is based on an end-to-end model and there is no such model for Quran recitation processing, we will discuss it in the MSA since there are some researchers who used end-to-end deep learning on modern standard Arabic. A comparison of these works is shown in Table 2.

Hussein et al. [29] proposed an End-to-End transformer-based Arabic Automatic Speech Recognition (ASR) model with a multitask objective function of Connectionist Temporal Classification (CTC)/Attention, where long short-term memory (LSTM) and transformer-based language model (TLM) were the two kinds of language models utilized in this work. The proposed model was compared to the previous approaches for Modern Standard Arabic (MSA) recognition task using Multi-Genre Broadcast 2 (MGB2) [30] data and for the Dialectal Arabic recognition task using MGB3 [31] and MGB5 [32] data. While the conventional word error rate (WER) was used to evaluate the model results for the first task, the multi-reference word error rate (MR-WER) and averaged WER (AV-WER) which adopted from MGB3 [31] and MGB5 [32] challenges, were used to evaluate the model results for the second task. 12.5%, 27.5%, 33.8% WER were achieved for the MGB2 [30], MGB3 [31], and MGB5 [32] challenges, respectively.

Ahmed et al. [33] introduced an end-to-end model based on Bidirectional Recurrent Neural Network with CTC objective function and a 15-gram language model as an Arabic speech-to-text transcription system. Also, a character-based decoder without a lexicon was used. This model was evaluated using 1200 hours corpus of the Aljazeera multi-Genre broadcast programs (MGP2) [30], where the WER was 12.03% for non-overlapping speech on the development set.

Belinkov et al. [34] analyzed the internal learned representations in an end-to-end ASR model for two languages (English and Arabic). Three datasets were used, Librispeech [35] and TED-LIUM [36] were used for English and the MGB-2 corpus [30] which has 1200 h from the Al Jazeera Arabic TV channel, was used for Arabic.

Alsayadi et al. [37] proposed end-to-end deep learning approaches to build a diacritized Arabic ASR. Two types of speech recognition approaches were used: The conventional ASR approach and the end-to-end ASR approach which consists of two models. The first model was built using Joint CTC attention based on the ESPnet toolkit [38] with an RNN-based language model, and the second model was built based on CNN-LSTM with the attention method using the Espresso toolkit [39] and with an external LM containing about 1.8 m words and 245k unique words. Training and testing of these models were done based on the Standard Arabic Single Speaker Corpus (SASSC), which contains 7 h of modern standard Arabic speech. The WER of 33.72%, 31.10%, and 28.48% were achieved for conventional ASR, the first end-to-end model, and the second end-to-end model, respectively.

**Table 2** Comparison between works related to end-to-end approaches for Arabic ASR

| Work | Dataset | Model | Language Model | WER |
|---|---|---|---|---|
| [29] | MGB2, MGB3, and MGB5 | transformer-based with a multitask objective function of CTC/Attention | LSTM and transformer-based | MGB2: 12.5%<br>MGB3: 27.5%<br>MGB5: 33.8% |
| [33] | MGP2 | Bidirectional Recurrent Neural Network based with CTC objective function | 15-gram | 12.03% |
| [37] | SASSC | Conventional ASR | 3-gram | 33.72%, |
| | | Joint CTC attention | RNN based | 31.10% |
| | | CNN-LSTM with attention | external LM | 28.48% |

# 3. Dataset

The dataset used in this work is the Ar-DAD dataset [40] which is a large dataset of Arabic-based audio clips containing 15810 clips of 30 popular reciters reading 37 chapters from the Holy Quran in addition to 397 audio clips for 12 imitators of the top reciters and two plain text files that contain the same chapters' textual content read by the reciters with and without vocalization (vowelization).

The audio samples, which are 10 seconds long on average and have a sampling rate of 44.1 kHz, 16-bit depth, and stereo channels, are shared in the WAV format.

The dataset was split as 80% for training, 10% for testing, and 10% for validation, where 12648 clips, 1581 clips and 1581 clips, were selected randomly and used for training, testing, and validation, respectively.

We noticed that the dataset contains all speeds of recitations mentioned before. In addition, the majority of the first verse samples transcripts of each chapter contain the sentence "بِسْمِ اللَّـهِ الرَّحْمَـٰنِ الرَّحِيمِ" while the corresponding audio clips do not contain the pronunciation of this sentence, so we removed this sentence from all transcripts because it would cause a problem in training the model since the number of these samples is about 1100 samples out of 15810 samples.

# 4. Methodology

The proposed solution consists of two main components: a CNN-Bidirectional GRU encoder and a character-based decoder. The encoder maps the input vector of features to a latent representation. The decoder takes the latent representation and generates one prediction at a time. CTC is the objective function used to train the encoder. The next subsections discuss them in detail.

## 4.1 Encoder

The encoder is CNN-BiGRU. The reason behind using CNN as the first layer is that the ASR performance can be improved by applying convolutions in frequency and time domains to spectral input features [41]–[43]. In addition, using Bidirectional RNNs in speech recognition provides better context utilization, both forward and backward, to accurately predict words [44].

The input of the encoder is the normalized spectrogram of audio clips. Each audio clip is a time series of length $T$ with a vector of audio features for each time slice. The input vectors $V_1, V_2, \ldots, V_T$ , are prepared by the 2D convolution layers (time and frequency domains) and then the CNN output is fed as input to Bidirectional GRUs. The output probabilities of the encoder are maximized using the CTC loss function.

### 4.1.1 Convolution Neural Network (CNN)

Convolutional Neural Networks (CNNs) are a type of deep learning architecture widely used for image classification and recognition tasks [45]. CNNs are designed to automatically learn the features of the input data and make predictions based on those features. The input data is typically processed in a series of convolutional layers, activation functions, and pooling layers.

In a convolutional layer, a set of kernels, also known as filters, slide over the input data, computing dot products between the input data and the weights of the kernels. The dot products are then used to produce feature maps, which are fed into the activation functions to introduce non-linearity to the model. The size of the kernels, the stride (the step size at which they move over the input data), and the padding (the addition of zeros around the input data to control the size of the output feature maps), are all hyperparameters that can be optimized for the specific task and input data.

2D Convolutional Neural Networks (2D-CNNs) are a specific type of CNN that operate on 2D input data, such as an image. In a 2D-CNN, the convolutional layers perform 2D convolutions, using 2D kernels to scan the input data and extract local features. The hierarchical representations built by the multiple convolutional and pooling layers allow 2D-CNNs to learn increasingly complex and discriminative features, making them well-suited for tasks such as object recognition and segmentation[46]. In addition, 2D-CNNs can be a powerful tool for speech recognition tasks, as they allow for the automatic extraction of relevant features from the spectrogram of an audio signal. By combining 2D-CNNs with other deep learning architectures, such as RNNs, end-to-end speech recognition systems can be created that can handle variable-length inputs and model the complex relationships between speech sounds [47].

### 4.1.2 Bidirectional Recurrent Neural Networks (Bi-RNNs)

A Bi-RNN is a combination of two RNNs that process two sequences, one in a forward direction and one in a backward direction, to benefit from the information of the past and the future and to compute the likelihood of the output character $c$ at a given time input $X_t$, depending on the previously hidden state $h_{t-1}$, the current input $X_t$ and the next hidden state $h_{t+1}$ [44].

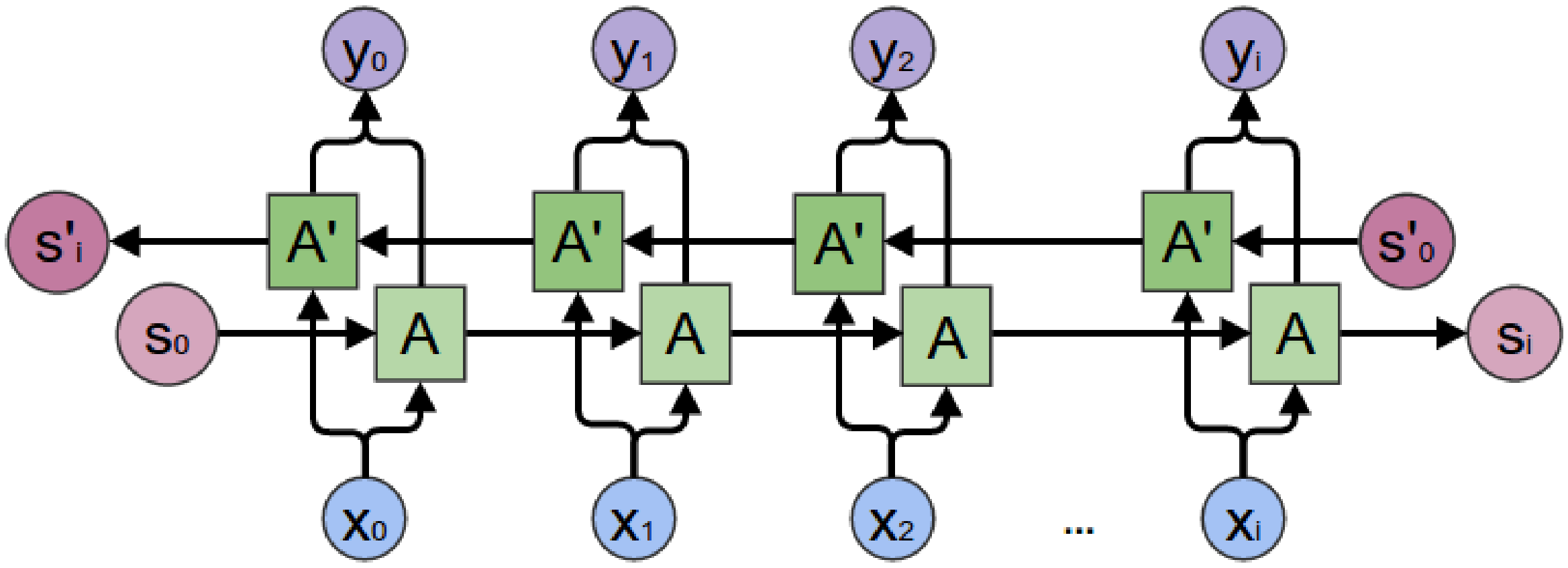

**Fig. 4** Bidirectional Recurrent Neural Network [48]

As shown in Figure 4, there is an additional-hidden layer for each Bi-RNN layer to accommodate the backward training process, where the forward and backward hidden states are updated at a given time $t$ as follows:

$$A_t(Forward) = f\left(X_t * W_{XA}^{forward} + A_{t-1}(Forward)\right) * W_{AA}^{forward} + b_A^{forward} \quad (1)$$

$$A_t(Backward) = f\left(X_t * W_{XA}^{backward} + A_{t+1}(Backward)\right) * W_{AA}^{backword} + b_A^{backward} \quad (2)$$

Where $b$ is the bias, $W$ is the weight matrix and $f$ is the activation function. And the hidden state is:

$$h_t = A_t(Forward) + A_t(Backward) \quad (3)$$

A Bi-RNN's network block can be vanilla RNNs that suffer from vanishing gradient [49] and exploding gradient problems [50], Gated Recurrent Units (GRU), or Long Short-Term Memory (LSTM). GRU and LSTM architectures are the most used when using long RNNs, because they handled vanishing and exploding gradient problems and are capable of learning long-term dependencies [51], [52]. In this work, we used GRU architecture because it has fewer parameters than LSTM and is faster to train (a GRU has two gates, reset and update gates, whereas an LSTM has three gates, input, output and forget gates). Also, it was discovered that the performance of GRU and LSTM were comparable for some tasks involving speech signal modeling and natural language processing [53], [54].

Figure 5 shows the architecture of a GRU cell and the equations to calculate the value of gates.

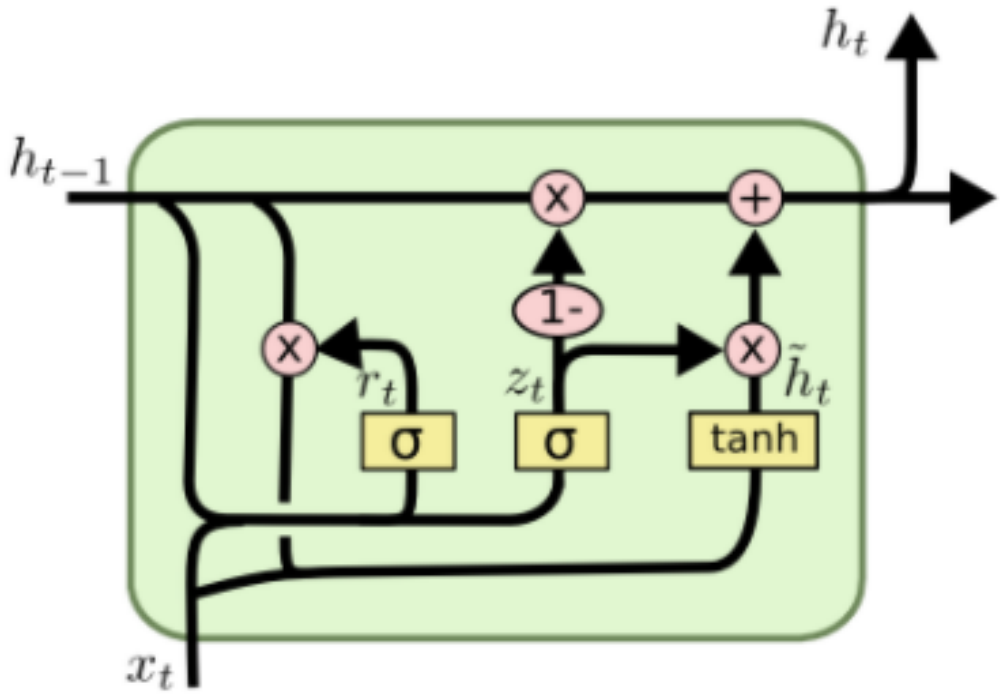

$$z_t = \sigma\left(W_z \cdot [h_{t-1}, x_t]\right)$$
$$r_t = \sigma\left(W_r \cdot [h_{t-1}, x_t]\right)$$
$$\tilde{h}_t = \tanh\left(W \cdot [r_t * h_{t-1}, x_t]\right)$$
$$h_t = (1 - z_t) * h_{t-1} + z_t * \tilde{h}_t$$

**Fig. 5** GRU architecture and equations [55]

Where $z$ and $r$ represent the update and reset gates respectively, $\sigma$ is the sigmoid function, $X_t$ is the current input, $W$ is the matrix of weights, ĥ is the current memory content and $h$ is the final memory at the current time step.

The Bi-RNNs layers are followed by a fully connected layer and an output layer which uses the softmax function to calculate the probability distribution over characters as follows:

$$p(c = k|x) = \frac{\exp(w_k^L \,.\, h_t^{L-1})}{\sum_j \exp(w_j^L \,.\, h_t^{L-1})} \quad (4)$$

$L$ represents the output layer, so $h^{L-1}$ is the hidden representation of the previous layer.

### 4.1.3 Connectionist Temporal Classification (CTC)

Connectionist Temporal Classification (CTC) [56] is the loss function used to train the model. CTC is an output and scoring function that addresses sequence problems when the alignment between the input and the output is not known, so it is applied in applications like speech and handwriting recognition.

For each time step $t$ and a single input sequence $X$ of length $T$, the encoder gives a distribution over the vocabulary, $p_t(c|X)$, then CTC computes the probability for a single sequence $C$ of length $T$ as follows:

$$P(C|X) = \prod_{t=1}^{T} p(c_t|X) \tag{5}$$

The same word can be represented by several different sequences, so finding the most likely sequence is done by summing over the probability of these sequences:

$$P(S|X) = \sum P(C|X) \tag{6}$$

Finally, CTC loss, which is the negative log probability of all valid sequences, is calculated using the dynamic programming algorithm which speeds up the calculation. In addition, to calculate its derivative, which utilizes the backpropagation through time algorithm to update the encoder's parameters. CTC loss function:

$$L_{CTC} = -\log P(S|X) \tag{7}$$

### 4.2 Decoder

In this paper, a character-level decoder was used because this type of decoder has several advantages over word-level decoders. One of the main advantages is that character-level models are more robust to out-of-vocabulary (OOV) words and variations in pronunciation [57]–[59]. Since the model is trained to predict individual characters, it is able to handle words that it has not seen before by predicting the individual characters that make up the word. This is particularly useful in speech recognition, where there may be many rare or unknown words. Character-level decoders also tend to be more computationally efficient than word-level decoders [60]. Since the model is only predicting individual characters, it does not need to search through a large vocabulary to find the most likely word. This can make decoding faster and more efficient.

In general, decoders are used to find the proper output for a given input by solving the following equation:

$$S^* = argmax_S\, p(S|X) \tag{8}$$

Greedy algorithms are used to solve this problem by using the output that is most likely at each time step. However, these algorithms have a big problem, which is overlooking the possibility that a single output sentence could have a variety of alignment forms [57]. For that, we used the CTC beam search decoder that sums the probabilities of each sentence to produce the best result. The CTC loss function is used to train the model, and the beam search decoder is used to generate the final output sequence. The beam search decoder works by maintaining a fixed number of top-scoring sequences (the "beam") at each decoding step, rather than considering all possible next steps. This reduces the search space and allows for faster decoding while still maintaining good accuracy. Additionally, the CTC loss function allows the decoder to be robust to variations in the timing of the input, making it well-suited for speech recognition tasks. Overall, the CTC beam search decoder is an efficient and effective method for decoding sequences in speech recognition and other sequence-to-sequence tasks [61].

## 5. Experiment Setup

Several speech recognition models that have shown highly accurate results were proposed in recent years. One such model is Deep Speech 2 [62]. Deep Speech 2 was developed by Baidu Research, it uses a combination of convolutional neural networks (CNNs) and recurrent neural networks (RNNs) to transcribe speech to text, as shown in Figure 6. The model starts with a convolutional layer that extracts features from the spectrogram, followed by several more convolutional layers that extract higher-level features. The output of these layers is then passed through a stack of bidirectional Long Short-Term Memory (LSTM) or Gated Recurrent Units (GRU) layers to process the sequential information in the speech, then a linear layer maps the output to the final text transcript. The training uses a Connectionist Temporal Classification (CTC) loss function that allows the model to align the output labels to the input speech, regardless of the length mismatch between the two.

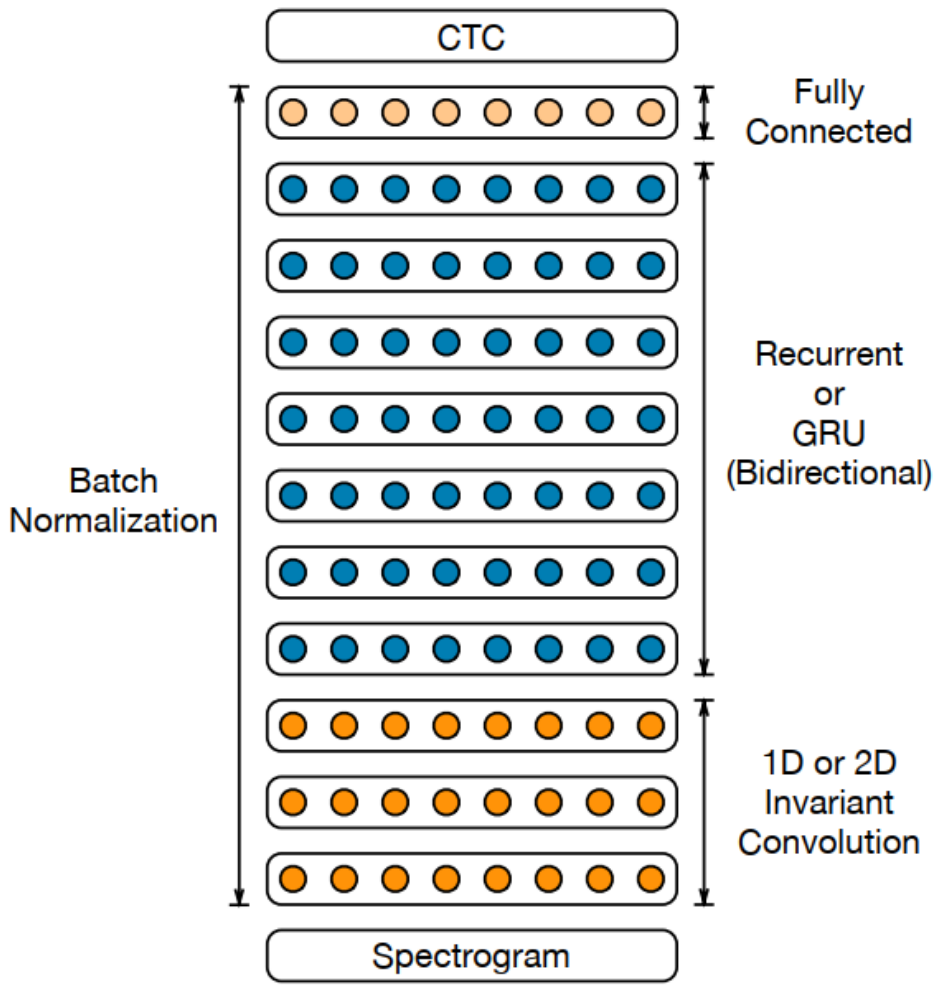

**Fig. 6** Architecture of Deep Speech 2 [62]

The architecture used in this work is shown in Figure 7. Our model consists of two 2D convolution layers with a kernel size of (11, 41) and a stride size of (2, 2) for the first layer, a kernel size of (11, 21) and a stride size of (1, 2) for the second layer, and 32 filters each. CNN layers are followed by 5 bidirectional GRU layers with 512 units for each layer and dropout layers of 0.5 rate. ReLU activation layers, and Batch normalization layers were also used. Batch normalization helps stabilize the training of the model and reduces the chances of overfitting by normalizing the activations of each layer to have zero mean and unit variance [63]. Rectified Linear Unit (ReLU) layers are a type of activation function that introduce non-linearity to the model, to help the model to learn more complex representations of the input data [64]. Then finally comes a dense layer containing 1024 neurons allows the model to learn interactions between different features, followed by a classification layer consisting of 46 neurons, to classify the current input into the blank symbol used in the CTC algorithm or the alphabet used in this work, which illustrated in Table 3.

**Table 3** The alphabet used in this work

| | | | | | | | | | |
|---|---|---|---|---|---|---|---|---|---|
| ح | ج | ث | ت | ب | آ | ا | أ | إ | ء |
| ط | ض | ص | ش | س | ز | ر | ذ | د | خ |
| ه | ن | م | ل | ك | ق | ف | غ | ع | ظ |
| ◌ً | ◌َ | ◌ّ | ◌ْ | ة | ئ | ى | ي | ؤ | و |
| | | | | 3 | Space | ◌ٌ | ◌ُ | ◌ٍ | ◌ِ |

Adam optimizer with 1e-4 learning rate and CTC loss function were used to train the model and was implemented using the TensorFlow library in Python and trained on the Google Colab platform[7] which provides access to an NVIDIA Tesla T4 GPU with 16 GB of memory.

[7] **Colaboratory**: https://colab.research.google.com/

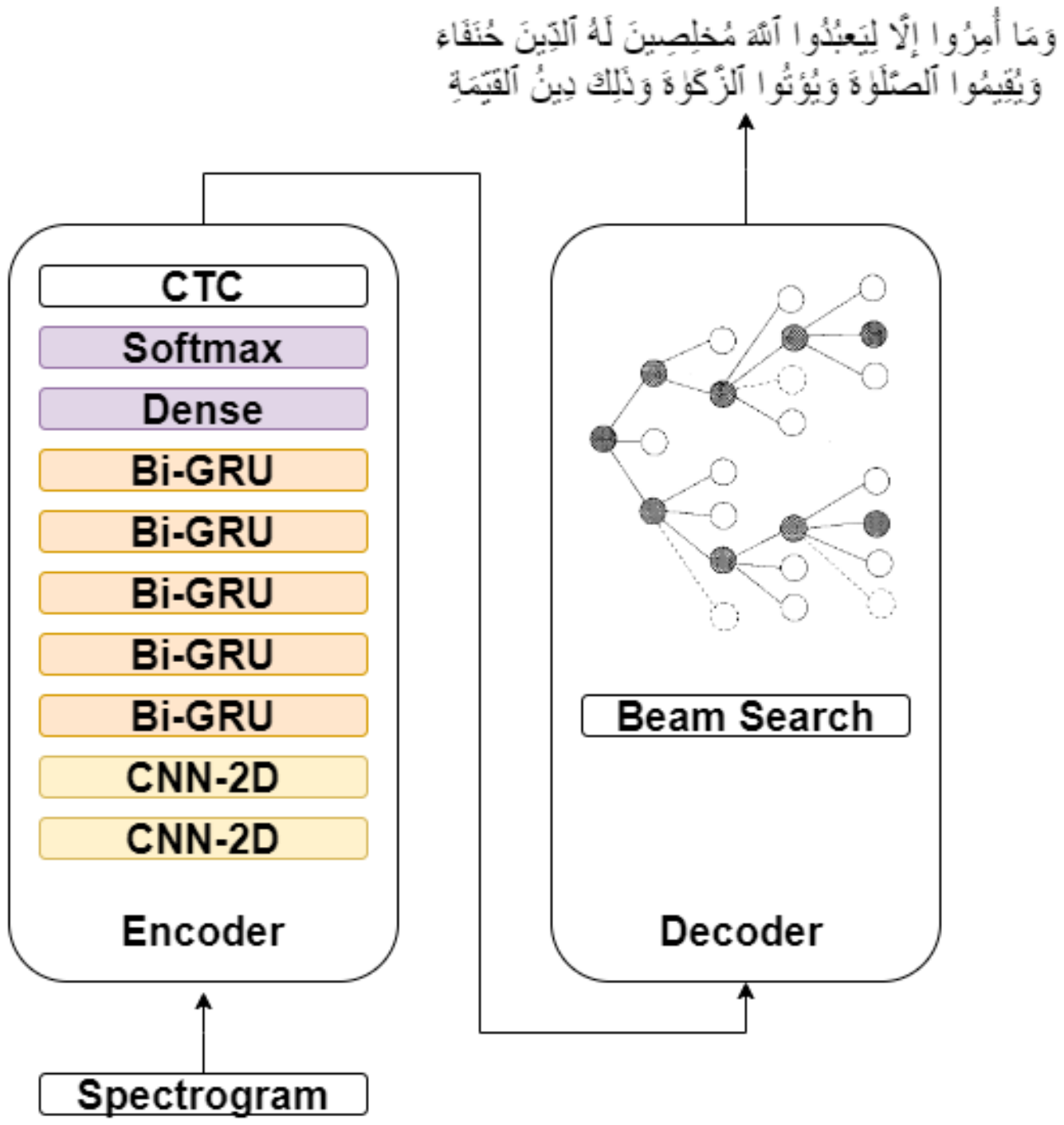

**Fig. 7** Our Proposed Solution Schema

# 6. Result & Discussion

As the Ar-DAD dataset includes recitations by a large number of readers at different speeds, we performed many experiments to choose the best spectrogram extraction parameters that can be suitable for all speeds. Choosing appropriate parameters, values of frame length, and hop size, in particular, is important to balance the trade-off between time and frequency resolution when Fourier Transform is applied. In practice, the choice of these parameters depends on the application and the characteristics of the signal being analyzed, and it's an iterative process. The detailed experiments are shown in Table 4.

We used the most common and widely used metrics for measuring the performance of speech recognition models, the Character Error Rate (CER), and the Word Error Rate (WER) metric. WER of 8.34% and CER of 2.42% were the best results we achieved.

The formula for Word Error Rate (WER) is:

$$WER = (S + D + I) / N \tag{9}$$

The formula for Character Error Rate (CER) is:

$$CER = (S + D + I) / M \tag{10}$$

Where: S, D and I are the number of substitutions, deletions, and insertions required to change the recognized transcript into the reference transcript. N is the total number of words, and M is the total number of characters in the reference transcript.

**Table 4** Experiments conducted to select feature extraction parameters

| Experiment Number | FFT | Hop Size | Test set | |
|---|---|---|---|---|
| | | | WER | CER |
| 1 | 512 | 256 | 9.02% | 2.64% |
| 2 | | 384 | 8.7% | 2.45% |
| 3 | 800 | 400 | **8.34%** | 2.77% |
| 4 | | 600 | 8.51% | **2.42%** |
| 5 | 1024 | 512 | 10.7% | 3.3% |
| 6 | | 768 | 11.5% | 3.2% |

As for comparing our results with the results of previous work, unfortunately, all the datasets used before are private and this made it difficult for us to compare with others. Therefore, using a public dataset in this research will solve the problem in the future so that researchers in this field can compare their work with us. However, as we have shown before, the vast majority of the previous works consider the problem as a classification task to detect the mispronunciation of Tajweed rules and some verses by samples containing wrong and correct pronunciation. As for the few works that proposed solutions based on traditional ASR, as mentioned in the literature review, all of them used data containing a few chapters of the Quran at the same speed and a few verses by a small number of readers who recited those verses. So, their results cannot be compared with our results as the dataset we used includes about 37 chapters of the Holy Quran in addition to a large number of reciters, which reaches 30 reciters, who recited these chapters at different speeds and with different applications of Tajweed rules. Table 5 shows a comparison between the real text and the text predicted by the proposed model for some verses from the Ar-DAD dataset.

**Table 5** Some samples that show the performance of the model

| Actual | Predicted |
|---|---|
| إِنَّ الَّذِينَ آمَنُوا وَعَمِلُوا الصَّالِحَاتِ أُولَئِكَ هُمْ خَيْرُ الْبَرِيَّةِ | إِنَّ الَّذِينَ آمَنُوا وَعَمِلُوا الصَّالِحَاتِ أُولَئِكَ هُمْ خَيْرُ الْبَرِيَّةِ |
| إِنَّ الَّذِينَ فَتَنُوا الْمُؤْمِنِينَ وَالْمُؤْمِنَاتِ ثُمَّ لَمْ يَتُوبُوا فَلَهُمْ عَذَابُ جَهَنَّمَ وَلَهُمْ عَذَابُ الْحَرِيقِ | إِنَّ الَّذِينَ فَتَنُوا الْمُؤْمِنِينَ وَالْمُؤْمِنَاتِ ثُمَّ لَمْ يَتُوبُوا فَلَهُمْ عَذَابُ جَهَنَّمَ وَلَهُمْ عَذَابُ الْحَرِيدِ |
| فَأَمَّا الْإِنْسَانُ إِذَا مَا ابْتَلَاهُ رَبُّهُ فَأَكْرَمَهُ وَنَعَّمَهُ فَيَقُولُ رَبِّي أَكْرَمَنِ | فَأَمَّا الْإِنْسَانُ إِذَا مَا ابْتَلَاهُ رَبُّهُ فَأَكْرَمَهُ وَنَعَّمَهُ فَيَقُولُ رَبِّي أَكْرَمَ |
| وَجِيءَ يَوْمَئِذٍ بِجَهَنَّمَ يَوْمَئِذٍ يَتَذَكَّرُ الْإِنْسَانُ وَأَنَّى لَهُ الذِّكْرَى | وَجِيءَ يَوْمَئِذٍ بِجَهَنَّمَ يَوْمَئِذٍ يَتَذَكَّرُ الْإِنْسَانُ وَأَنَّى لَهُ الذِّكْرَى |
| لَمْ يَكُنِ الَّذِينَ كَفَرُوا مِنْ أَهْلِ الْكِتَابِ وَالْمُشْرِكِينَ مُنْفَكِّينَ حَتَّى تَأْتِيَهُمُ الْبَيِّنَةُ | لَمْ يَكُنِ الَّذِينَ كَفَرُوا مِنْ أَهْلِ الْكِتَابِ وَالْمُشْرِكِينَ مُنْفَكِّينَ حَتَّى تَأْتِيَهُمُ الْبَيِّنَةُ |
| وَمَا أُمِرُوا إِلَّا لِيَعْبُدُوا اللَّهَ مُخْلِصِينَ لَهُ الدِّينَ حُنَفَاءَ وَيُقِيمُوا الصَّلَاةَ وَيُؤْتُوا الزَّكَاةَ وَذَلِكَ دِينُ الْقَيِّمَةِ | وَمَا أُمِرُوا إِلَّا لِيَعْبُدُوا الّهَ مُخْلِصِينَ لَهُ الدِّينَ حُنَفَاءِنَ وَيُقِيمُوا الصَّلَاةَ وَيُؤْتُوا الزَّكَاةَ وَذَلِكَ دِينُ الْقَيِّمَةِ |
| إِنَّ الَّذِينَ كَفَرُوا مِنْ أَهْلِ الْكِتَابِ وَالْمُشْرِكِينَ فِي نَارِ جَهَنَّمَ خَالِدِينَ فِيهَا أُولَئِكَ هُمْ شَرُّ الْبَرِيَّةِ | إِنَّ الَّذِينَ كَفَرُوا مِنْ أَهْلِ الْكِتَابِ وَالْمُشْرِكِينَ فِي نَارِ جَهَنَّمَ خَالِدِينَ فِيهَا أُولَئِكَ هُمْ شَرُّ الْبَرِيَّةِ |

# 7. Conclusion

In conclusion, this research presents a novel end-to-end deep learning model for recognizing Holy Quran recitation. Moreover, our proposed model provides the ability to give feedback to users about error type and location so that users can have a better experience in correcting their errors in their learning journey. The proposed solution consists of two main components: A CNN-Bidirectional GRU encoder uses the CTC loss function and a character-based decoder. Using this end-to-end model allows us to get rid of alignment tools, so reduce needed efforts, and improve performance. Our proposed model has been evaluated on a recently published public data set (Ar-DAD), which contains about 37 chapters recited by 30 reciters. Ar-DAD dataset used to train a model for recognizing Quran recitation has limitations. It only includes men reciters, making it less reliable for recognizing recitations by women and children. Additionally, it only contains samples from one recitation form, while there are ten approved forms, which may cause the model to incorrectly recognize recitations with different forms. The model's performance was evaluated using word error rate (WER) and character error rate (CER) as metrics, with the best results obtained being 8.34% WER and 2.42% CER. These results demonstrate the effectiveness of the proposed model in recognizing the recitation of the Holy Quran that outperforms previous related works. We hope that this paper will provide a baseline for fair comparisons in this task, as it is based on a publicly available dataset that can be used by all researchers.

# 8. Declarations

## Authors' contributions

AAH performed the literature review, conducted the experiments, and wrote the manuscript.
KAJ took on a supervisory role and made a contribution to the conception and analysis of the work.
All authors read and approved the final manuscript.

## Funding

The authors declare that they have no funding.

## Data Availability

The data set used in this work is available at:

https://data.mendeley.com/datasets/3kndp5vs6b/3

## Ethics approval and consent to participate

The authors Ethics approval and consent to participate.

## Consent for publication

The authors consent for publication.

## Conflicts of Interest

The authors declare that they have no competing interests.